\documentclass[useAMS,usenatbib]{mn2e}
\input epsf.sty

\title[Intranight variability in quasars]{The nature of the intranight 
variability of radio-quiet quasars} 

\author[B. Czerny, A. Siemiginowska, A. Janiuk, A. C. Gupta] 
{B. Czerny$^{1}$\thanks{E-mail:
bcz@camk.edu.pl}, A. Siemiginowska$^{2}$, A. Janiuk$^{1}$,
A.C. Gupta$^{3}$ \\
$^{1}$ Nicolaus Copernicus Astronomical Center, Bartycka 18,
            00-716 Warsaw, Poland\\
$^{2}$ Harvard Smithsonian Center for Astrophysics, 60 Garden St, Cambridge, MA 02138 \\
$^{3}$ Aryabhatta Research Institute of Observational Sciences
(ARIES), Manora Peak, Nainital - 263129, India \\
}

\begin{document}

\pagerange{\pageref{firstpage}--\pageref{lastpage}} \pubyear{2007}

\maketitle

\label{firstpage}

\begin{abstract}

We select a sample of 10 radio-quiet quasars with confirmed intranight
optical variability and with available X-ray data. We compare the 
variability properties and the broad band spectral constraints to the 
predictions of intranight variability
by three models: (i) irradiation of an accretion disk by a variable X-ray flux 
(ii) an accretion disk instability (iii) the presence of a weak blazar component. 
We concluded that the third model, e.g. the blazar component model, is the 
most promising if we adopt a cannonball model for the jet variable emission. 
In this case, the probability of detecting the intranight variability 
is within 20-80\%, depending on the ratio of the disk to the jet optical 
luminosity. Variable X-ray irradiation mechanism is also possible but only 
under additional requirement: either the source should have a very narrow
H$\beta$ line or occasional extremely strong flares should appear at very large 
disk radii. 

\end{abstract}


\section{Introduction}
\label{sec:intro}

Active galactic nuclei (AGN) are variable on all timescales and in all
energy bands. This variability can be used to improve understanding of
the accretion process onto a massive central black hole that powers
the nuclear activity, 
because due to the very short timescales we can probe the conditions 
very close to the center. The variability mechanisms can be tested using 
the properties of the X-ray emission, as well as by the analysis of the 
optical data.

The most puzzling variations are those happening on the shortest timescales, 
of a fraction of a day. Variations in flux of a few tenth of magnitude 
in time scale tens of minutes to tens of hours (less than a day) are often 
known as microvariability or intranight variability or intraday variability 
(Wagner \& Witzel 1995). 
The fast intraday variability was first discovered in the X-ray band 
(Mushotzky, Holt \& Serlemitsos 1978) and with improvement of X-ray detectors 
were found to be a typical property of radio-quiet AGN (e.g. McHardy \& Czerny
1987; Lawrence et al. 1987; Uttley, McHardy \& Papadakis 2002; McHardy
et al. 2004), although the level of variability strongly depends on
the black hole mass (Markowitz \& Edelson 2004). A fast intraday
variability in the radio band was later discovered in
radio-loud objects (Witzel et al. 1986; Heeschen et al. 1987), and now
it is considered to be a common phenomenon in such objects at
centimeter wavelengths (e.g. Lovell et al. 2003, Kraus et
al. 2003). Later similar fast variations were found in the optical
band, also in radio-loud objects, (Miller, Carini \& Goodrich 1989; 
Heidt \& Wagner 1996; Gupta et al. 2008; and references therein); 
in that case it is more appropriate to call it intranight variability.
Finally, extensive optical observational campaigns of radio-quiet 
objects shown an occasional presence of intranight variability 
(Gopal-Krishna et al. 2000, 2003; Stalin et al. 2004, 2005; Gupta \& Joshi 
2005).

Intraday variability observed in radio-loud objects in low frequency
radio observations is either caused by the interstellar scintillation,
as supported by the intra-hour variability (Kedziora-Chudczer et al. 1996; 
see also Kedziora-Chudczer 2006; Raiteri et al. 2006) or it is intrinsic 
and related to jet instabilities, as indicated by a match of its amplitude 
to the longer trends for 3C 279 (Kartaltepe \& Balonek 2007). Actually, both 
effects may be at work in the radio-loud sources. The nature of the observed 
intranight variability of radio-quiet objects is unknown.

The variability mechanisms in the radio-loud objects are widely
believed to be connected to the conditions in the jet. However, it is 
still unclear, if in the radio-quiet objects the nature of intranight 
variability (INV) is different, or whether the faint, variable jet is here 
a dominant INV component as well.

In the present paper we consider three possible mechanisms 
for an intranight optical variability of radio-quiet quasars: reprocessing 
of the variable X-ray emission by an accretion disk, intrinsic variability 
of the accretion disk, and the presence of a weak blazar component. 
To better distinguish between these three scenarios, we make an attempt to 
correlate the optical and X-ray information. We confront the model predictions
with the observational data for a selected sample of 10 objects. 

In Section 2, we summarize the optical and X-ray data. In Section 3, we
describe our three models. In Section 4, we give the results and in Section 5, 
we discuss and conclude.

\section{Data}
\label{sec:data}

\begin{table*}
{\scriptsize
\noindent
\caption{Micro-variable Radio-Quiet Quasars with available X-ray data$^1$}
\begin{tabular}{lcccccccccc}
\hline \hline
Source  &  RA(2000.0) & DEC(2000.0) & z &  R &   V &   0.1-2.4 keV & $\Gamma_X$  & $\alpha_{ox}$ \\
\hline
PG 0026+129 & 00 29 13.7 & $+$13 16 05 & 0.142 & 1.08 & 15.41 & ~9.37 &	2.31 (ROSAT) & 1.30 \\
MKN 1014    & 01 59 50.1 & $+$00 23 41 & 0.163 & 2.12 & 15.69 & ~4.03 & 2.82 (ROSAT) & 1.52 \\
PG 1116+215 & 11 19 08.7 & $+$21 19 18 & 0.177 & 0.72 & 14.73 & 12.00 & 2.62 (ROSAT) & 1.43  \\
1750+507    & 17 51 16.7 & $+$50 45 39 &   0.3 & 5.01 & 15.40 & 11.73 & 3.03 (ROSAT) & 1.47 \\
AKN 120     & 05 16 11.4 & $-$00 08 59 & 0.032 & 1.03 & 14.10 & ~3.6$^3$ & 2.46 (XMM)& 1.44  \\
Q 1252+020  & 12 55 19.7 & $+$01 44 12 & 0.342 & 0.52 & 15.48 & ~5.95 & -            & 1.29 \\
\hline
upper limits &   &&& && flux at 1 keV & & $\alpha_{ox}$ \\
0824+098 &  08 27 40.1 &  09 42 10 & 2.928 & - & 18.3 &  $ < 8.83e-14$  && $> 1.35$ & \\
PG 0832+251 &  08 35 35.9 & 24 59 41 & 0.331 &   1.26 & 16.1 & $< 9.3e-14$ && $ > 1.61$ \\
PG 0043+039 & 00 45 47.2 &  04 10 24 & 0.385 &  - &  16.0 (B) & $<8.6e-16$ && $ > 2.3$ \\
\hline
\label{tab:xrays}
\end{tabular}

{\footnotesize $^1$ columns denote source name, RA (2000.0), DEC (2000.0),
redshift, radio loudness parameter, visual magnitude, X-ray flux in 1.e-14 ergs/s/cm$^2$ 
from Yuan et al. 1998, soft X-ray slope and broad-band optical/X-ray slope; $^2$ (0.5 - 2 keV) 
from Risaliti et al (2003); $^3$ from Vaughan et al (2004), in 1.e-11 erg/s/cm2 (2-10 keV) , 
with $\Gamma = 2.0$.  }}
\end{table*}

Our aim is to confront the 
intranight variability models and the observational data taking 
into account the broad band spectral constraints. We focus on X-ray and 
optical bands because in many models they are strongly related. 
Therefore, we collect a sample of radio-quiet quasars
with a confirmed 
intranight variability in the optical band that were also
observed in the X-ray band.

\subsection{Optical Data}

A sample of 117 objects with reliable studies of 
intranight variability has
been compiled by Carini et al. (2007). Since Seyfert galaxies are
strongly contaminated by the starlight, we removed from the sample all
sources with $M_B > -24.0 $. The remaining sample of 96 objects
contains 22 sources with detected  
intranight variability (for the remaining monitored 74 sources 
intranight variability was not detected).  We
supplement this sample with 1 source with detected 
intranight variability from Stalin et al. (2005) that is missing 
in Carini et al. (2007) compilation. We removed PKS 1103-006 and 
B2 1225+317 since they are definitively radio-loud objects (R=260 based 
on NED).

\subsection{X-rays}

There have been no systematic X-ray studies of the sources with the
observed intranight optical variability. We searched available
published data and X-ray data archives to obtain the X-ray information
for the variable sources. Note that we only consider X-ray properties
that are non-simultaneous with the optical data and have no
information on correlation between the optical and X-ray variability
for these sources.  While it is important to understand the overall
variability considering simultaneously both optical and X-ray bands at
this time we can only discuss model constraints based on the currently
available information. Systematic studies are needed in the future to
describe the characteristics of the X-ray variability, for example is
the source amplitude vary together with the spectral slope, or the
slope remains constant.


ROSAT PSPC X-ray data for 24 sources from the entire Carini et al. 
(2007) sample (5 variable and 19 non-variable) were published by 
Yuan et al. (1998). We supplement these X-ray data with observations from 
{\it Chandra} X-ray Observatory (1422+424; Risaliti et al. 2003) and
XMM-{\it Newton} (Ark 120, Vaughan et al. 2004), and with the
constraints for 3 more objects (PG0043+039, 0824+098 and 1422+424)
derived from the X-ray archival data (see Appendix).  

A summary of the X-ray properties for intranight variable radio-quiet
sources is presented in Table 1.  The soft X-ray flux has been
measured for 7 sources detected by ROSAT, while 3 other sources have
only upper limits calculated from the archival data.

The ROSAT PSPC slopes were measured for 4 of the variable sources, and
the average slope of these four sources ($\Gamma = 2.7 \pm 0.3$) is steeper
than the average slope in the non-variable sources in this sample
($\Gamma = 2.4 \pm 0.5$), but the dispersion in both cases is very large.
This is very interesting since the radio-loud quasars have on average
flatter (lower photon index) soft X-ray slopes than the radio-quiet
quasars (e.g. composites of Laor et al. 1997; SDSS by Richards et al
2006).
The median values of the soft X-ray slope for all the objects in the
sample is $\Gamma=2.51$. This is smaller than the average slope based
on just the four ROSAT slopes.

\subsection{Sample of variable sources}

Our sample of sources with a confirmed optical microvariability and 
with available X-ray constraints contains 10 objects. Their overall
properties are given in Table~\ref{tab:xrays}.

We determine the 
broad band spectral index for these sources, $\alpha_{ox}$, measured
between 2500 \AA~ and 2 keV (see e.g. Shen et al. 2006).  We assume
the optical slope, $p$, of 0.5 (in $F_{\nu} \propto \nu^{-p}$
convention) to obtain the 2500 \AA~ flux from the measured V magnitude
and we use a measured photon index and a flux in the X-ray band. The
calculated $\alpha_{ox}$ for each source are shown in Table~1.  For
the sources with the upper limit to the X-ray flux we calculate a
lower limit on $\alpha_{ox}$.

The median value of $\alpha_{ox}$ is 1.47 which is rather large (see
e.g.  Bechtold et al. 2003), in contrast to the recent studies of
large quasar samples (Shen et al. 2006) indicating that objects with
microvariability are X-ray over-luminous for their optical flux by a
factor 2, taking into account that our objects are not extremely optically
bright (steeper slopes are expected for brighter objects, see e.g. Vasudevan \&
Fabian 2007).  On the other hand, 18 non-variable sources from Carini et
al. (2007) sample with ROSAT X-ray data available show similarly large
$\alpha_{ox}$ of 1.51.

The kinematic width of
the broad emission lines is known indicator of black hole mass.
But what is more, the equivalent width of H$\beta$ line 
may indicate on the presence of an additional spectral component.
Therefore, 
for objects contained in SDSS data base\footnote{http://www.sdss.org} we
estimate both the equivalent width of the H$\beta$ line, without the
decomposition into a narrow and a broad component, as well as the
kinematical width (FWHM). The results are given 
in Table~\ref{tab:mass}. 

The FWHM was used to determine the black
hole mass, following closely the approach given by Vestergard \&
Peterson (2006). The obtained values range from $10^8 M_{\odot}$ to
just above $10^9 M_{\odot}$, somewhat smaller than average quasar mass
of $10^9 M_{\odot}$ but well within the mass distribution shown by 
Vestergaard et al. (2008). We also estimated the bolometric luminosities
of the sources assuming $L = 9 \lambda L_{\lambda} (5100)$. 
The Eddington ratio for these
objects is about 0.7 (adopting $L_{Edd} = 1.36 \times 10^{38}
(M/M_{odot})$ erg s$^{-1}$). It seems surprisingly uniform but we cannot
draw too strong conclusion based on four objects.

We also determined the H$\beta$ line intensity since the low value of
EW of the line serves as an indicator of the dilution of the
underlying continuum. The measured
equivalent widths are comparable with the typical
values, like 62.4 \AA~ (also a single line fits) in the Bright Quasar
sample (Forster et al. 2001). 


\begin{table*}
\caption{H$\beta$ line properties and black hole mass determination
  for sources present in SDSS}
\begin{center}
\begin{tabular}{lccccc}     
\hline\hline       
source     & EW(H$\beta$)  & FWHM(H$\beta$)
& $\lambda L_{\lambda}(5100)$ & $\log M$ & $\log L/L_{Edd}$\\
  & Angstroem  & km s$^{-1}$ & $10^{44}$ erg s$^{-1}$ cm$^{-2}$ 
&  & \\
\hline
 MKN 1014    & $-$37$\pm$3 & 2230$\pm$110  & ~10.8 &  8.1 & $-$0.27 \\
 1422+424    & $-$82$\pm$4 & 3380$\pm$140  & ~54.7 &  8.8 & $-$0.28\\
 Q 1252+020  & $-$60$\pm$3 & 4100$\pm$140  & 101.0 &  9.1 & $-$0.32\\
 PG 0832+251 & $-$74$\pm$4 & 3380$\pm$~90  & ~51.8 &  8.8 & $-$0.29\\

\hline
\end{tabular}
\end{center}
\label{tab:mass}
\end{table*}

\section{Models}
\label{sec:models}

We consider three plausible mechanisms of the microvariability in
radio-quiet AGN. The first one is the X-ray irradiation of an
accretion disk: all radio-quiet AGN are known to be strongly variable
in X-ray band (e.g. Mushotzky, Done \& Pounds 1993; Uttley, McHardy \&
Papadakis 2002; Markowitz \& Edelson 2004) so the reprocessing of the
variable X-ray flux is likely to lead to some optical variability, and
the phenomenon is well known at longer timescales (Rokaki,
Collin-Souffrin \& Magnan 1993; Gaskell 2006). The second mechanism is
the accretion disk instability suggested in the context of
microvariability by Mangalam \& Wiita (1993). The third mechanism,
suggested by Stalin et al. (2005) is the presence of the weak blazar
component even in radio-quiet objects; evidence for a jet in such
objects was discussed by Blundell \& Rawlings (2001), Barvainis et
al. (2005) and Leipski et al. (2006).

For each of the mechanisms we formulate a numerical model which can be
confronted with the observational constraints in the optical and X-ray
band.


\subsection{X-ray irradiation of an accretion disk}
\label{sec:irrad}
 
In this scenario we consider the strongly variable X-ray 
emission from a hot plasma above an accretion disk. This X-ray 
emission can be partially intercepted by the disk and thermalized, 
thus leading to the variable optical/UV emission. 

\subsubsection{lightcurves}

The X-ray variability properties were studied in numerous papers
(e.g. McHardy et al. 2004; Markowitz \& Edelson 2004; for a recent
review see for example Uttley 2007). In particular, the power spectrum
density was obtained for several AGN, and the scaling of the
variability with the mass and the luminosity was recently discussed by
McHardy et al. (2006).

In order to create the representative X-ray lightcurve we
thus use the popular Timmer \&  Koenig (1995) approach which allows to 
recover the lightcurve from the known power density spectrum. 

For the purpose of the modelling we assume that the power spectrum of
a representative AGN can be described by a power law with two breaks,
although very detailed study performed for Akn 564 show that multiple
Lorentzian representation is more accurate (McHardy et al. 2007). 
However, for our purpose the simple power law description is satisfactory. 
Therefore, we parameterize the power spectrum density with the high 
frequency break, $f_h$, and low frequency break, $f_l$, assuming the 
slope 0 below $f_l$, slope 1 between $f_l$ and $f_h$, and
slope 2 above $f_h$. The existence of the low frequency break is not
well established in AGN (apart from the case of {\bf Akn 564}, McHardy et
al. 2007) but the analogy with galactic X-ray sources suggests it may
be there; it is not essential for the current modelling since we aim
here at modeling the short timescale phenomena. Here we simply assume
that $f_l = 0.01 f_h$. The level of the power spectrum on $power
\times f$ diagram is fixed in all models at 0.01 in dimensionless
units. The actual value is difficult to determine, since the slope of
the flat part is not exactly zero, e.g. for Akn 564 the value rises
towards longer frequencies reaching 0.02 at maximum (McHardy et
al. 2006). However, adopting higher value right at the frequency break
overpredicts the variability in 1 day timescale measured by Markowitz
et al. (2003).

The positions of the breaks and the normalization of the high
frequency tail of the PSD are certainly related to the mass and/or
luminosity of the object (Hayashida et al. 1998; Czerny et al. 2001; 
McHardy et al. 2006; Gierli\' nski et al. 2008; Liu \& Zhang 2008). 
We parameterize the high frequency break using one of the two 
relations found by McHardy et al. (2006):
\begin{equation}
\log f_h = - 2.10 \log M_6 + 0.98 \log L_{44} + 2.33~~ [{\rm
    day}]^{-1}
\label{eq:break1}
\end{equation}
where $log M_6$ is the black hole mass in $10^6 M_{\odot}$ units and
$L_{44}$ is the bolometric luminosity in units of $10^{44}$ erg
s$^{-1}$,
or directly from the line width of H$\beta$ line width 
\begin{equation}
\log f_h  = - 4.2 \log FWHM + 14.43~~ [{\rm day}]^{-1},
\label{eq:break2} 
\end{equation}
where $FWHM$ is in km s$^{-1}$. We use this parameterization of power
spectrum density to generate X-ray lightcurves.

\begin{figure}
\epsfxsize=8.8cm 
\epsfbox{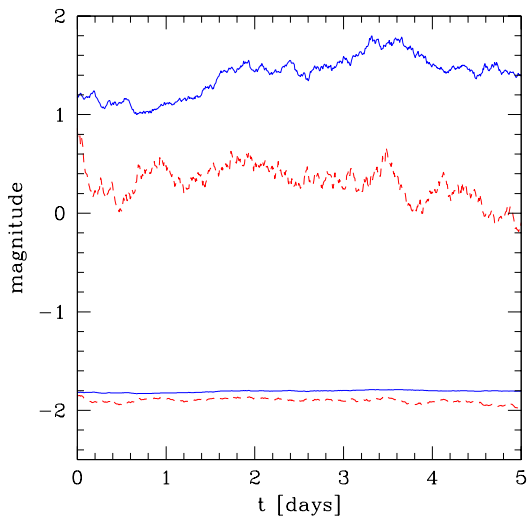}
\caption{Exemplary lightcurves (relative magnitudes) from X-ray irradiation
 model for two parameterization: X-ray coronal emission
(two upper curves) and the disk optical emission after reprocessing 
(two lower curves). 
Parameters:$M = 10^{8} M_{\odot}$, $L = 10^{45}$ erg s$^{-1}$
 (continuous line), $FWHM = 3000$ km s$^{-1}$ (dashed lines); 
$L_{\rm cor}/L_{\rm disk} = 0.1$. 
\label{fig:X_curves}}
\end{figure}

The optical variability in this scenario is caused by the
reprocessing of the variable X-ray emission by the disk. We
now assume that the X-ray variability is strongly localized,
i.e. we neglect the smearing effect caused by the reprocessing
which generally would lower the variability amplitude in the optical band.
An exemplary set of X-ray and optical lightcurves is shown in 
Fig.~\ref{fig:X_curves}. {\bf The incident flux is fully absorbed, 
i.e. we neglect the scattering effect (more accurate description should
include albedo of order of 0.2 - 0.5, depending on the ionization of
the disk surface)}. We also assume that many coronal flares
coexist, and the energy dissipation in flares follows the radial
profile of the energy dissipation in the disk. In such case we can
assume that the contribution of the X-ray reprocessed radiation is
wavelength-independent and the amplitude is determined by the
time-dependent corona to disk luminosity ratio. 

Disk lightcurve with the reprocessing effect shows fast
variability. For each lightcurve, we calculate the probability that a
fluctuation larger than 2 per cent is seen during the typical one
night observation (5 hours), where the fluctuation during a single
night, $\psi$, is determined as in Stalin et al. (2005)

\begin{eqnarray}
\psi &=& 100 \times \sqrt{(D_{\rm max} - D_{\rm min})^2 - 2 \sigma^2}= \nonumber \\
     & & -2.5 \times 100 \times \log(F_{\rm max}/F_{\rm min}),
\label{eq:psi} 
\end{eqnarray}
where $D_{\rm max}$ and  $D_{\rm min}$ are the observed magnitudes of
an object, and $F_{\rm max}$ and
$F_{\rm min}$ are the maximum and the minimum flux in the model. In the models we neglect 
the correction $\sigma^{2}$ for the observational error.

\subsubsection{broad band spectra models}

Since the average X-ray luminosity to the optical (disk) luminosity is
a free parameter of the model we construct the broad band spectrum
which allows to convert the corona/disk luminosity ratio to the
observational parameter $\alpha_{ox}$.

We assume that the local disk emission is well approximated by a black
body emission, and the total disk emission is given by an integral
over the disk surface.

The spectral shape of the coronal emission is not determined within
the frame of our model since we do not consider heating/cooling
effects for electrons. Therefore, we simply assume that the coronal
emission has the typical spectral slope of radio quiet quasars,
i.e. energy index $\alpha_E$ is 1.0 (i.e. photon index is equal 2.0).

\subsection{Accretion disk instability}
\label{sec:instab}

\subsubsection{Radiation pressure instability}
 
Optically thick accretion disks in AGN are radiation-pressure
dominated, unless the accretion rate is extremely low. The viscous and
thermal stability of such disks depend on the description of the
viscosity. Standard parametrization of Shakura \& Sunyaev (1973) of
the viscous torque by the total pressure ($\alpha P_{tot}$ assumption)
leads to disk instability (Pringle, Rees \& Pacholczyk 1973; Lightman
\& Eardley 1974; Shakura \& Sunyaev 1976) while the assumption that
the total torque scales with the gas pressure ($\alpha P_{gas}$
assumption) leads to stable solutions (Coroniti 1981). The disk
viscosity law determines the expected level of the disk intrinsic
variability. Numerical computations of the $\alpha P_{\rm tot}$ disk
models show violent periodic outburst with very large amplitudes, of a
few orders of magnitude (Szuszkiewicz \& Miller 1998; Li, Xue \& Lu
2007). Such gigantic outbursts are not seen in the observational data
for AGN or galactic sources. On the other hand, lower amplitude
outbursts are seen in GRS 1915+105 (for a review of the source
properties, see Fender \& Belloni 2004 and Janiuk, Czerny \&
Siemiginowska 2002 for applications of the radiation pressure instability
model).  This means that a viscosity law intermediate between $\alpha
P_{tot}$ and $\alpha P_{gas}$ is likely to represent well the disk
behaviour.  Various viscosity modifications were proposed in the past
(e.g. Taam \& Lin 1984; Szuszkiewicz 1990; Watarai \& Mineshige 2003).
Here we follow the recent work by Merloni \& Nayakshin (2006) and we 
adopt the geometric mean law, i.e. $\alpha (P_{gas}P_{tot})^{1/2}$.

\begin{figure}
\epsfxsize=8.8cm 
\epsfbox{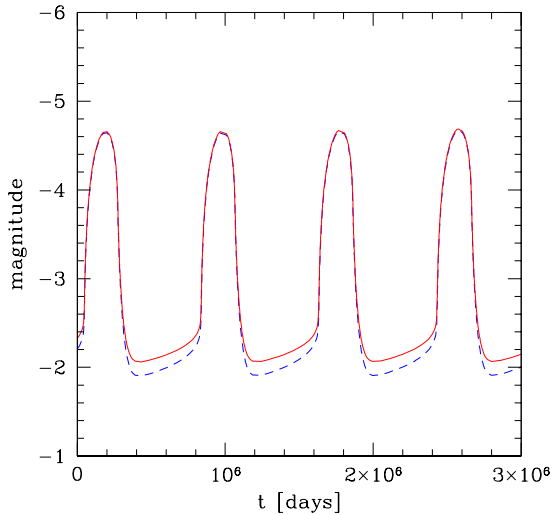}
\caption{The luminosity outbursts due to the accretion disk instability.
Parameters: $M_{BH} = 10^{8} M_{\odot}$, $\dot M = 0.8 M_{\odot}/yr$,
$\alpha_{\rm disk} = 0.01$. Dashed line shows the disk bolometric
luminosity and the continuous line shows the effect of the coronal irradiation.
\label{fig:outburst}}
\end{figure}

Computations of the time-dependent disk evolution under the influence
of radiation pressure instability were performed using the code of
Janiuk \& Czerny (2005) but with different viscosity law 
($\alpha (P_{gas}P_{tot})^{1/2}$ instead of $\alpha P_{gas}$).
Model predicts strong regular outbursts lasting thousands of years.
In this case no X-ray emission is predicted by the model if the option
for corona formation is switched off. Therefore, broad band constraints
cannot be used for this model. 

We can also include both the radiation
pressure instability and the effect of the corona formation. In this
case we use the same code, and we allow for the corona formation using
the Markoff chain formalism (King et al. 2004; Mayer \& Pringle
2006). The details of the computations are given in Janiuk \& Czerny
(2007). The basic outburst pattern does not change but an additional
flickering due to X-ray irradiation appears. 
An exemplary lightcurve with the corona effect included is
shown in Fig.~\ref{fig:outburst}.

\subsubsection{Magnetorotational instability}

\begin{figure}
\epsfxsize=8.8cm 
\epsfbox{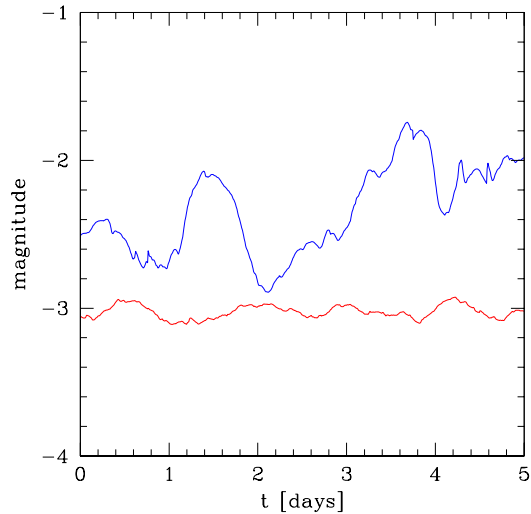}
\caption{The lightcurves expected from the local MRI developments at
  $10 R_{Schw}$ radius around a $10^7 M_{\odot}$ black hole for 36
  (upper blue curve) and 360 (lower red curve) coexisting magnetic cells.
\label{fig:MRI_36_360}}
\end{figure}

Magnetorotational instability (MRI), rediscovered by Balbus \& Hawley (1991)
in the context of accretion disks is the most likely source of the
disk viscosity. It roughly operates in the local dynamical timescale,
as confirmed by 2-D and 3-D simulations (for a review, see Balbus \&
Hawley 1998). 

Local development of the MRI can be roughly represented using the idea
of the Markoff chain (e.g. King et al. 2004, Mayer \& Pringle 2006, 
Janiuk \& Czerny 2007). The local magnetic field, $B$, is thus proportional to
the random number $u(k)$, generated in the following way
\begin{equation}
u(k+1) = - \alpha_1 u(k) + \epsilon_1,
\end{equation}
where $\epsilon_1$ is the random number with uniform distribution,
mean value 0 and variance equal 1, while $\alpha_1$ is a constant
describing the coupling in the system. King et al. (2004) assumed
$\epsilon_1 = -0.5$ and we will do the same throughout the paper.
The timestep is determined by the timescale of the magnetic field
development which is equal to $k_{mag} t_K$, where $t_K$ is the local
Keplerian period and $k_{mag}$ is a constant expected to be between 1
and 10. We will assume $k_{mag} = 1$ in order to create most favorable
conditions for the INV. The local disk dissipation is thus
proportional to $B^2 \propto u(k)^2$. 

The strength of the variability is determined by the number of
magnetic cells. Two examples of the lightcurves assuming the
coexistence of 36 and 360 magnetic cells are shown in
Fig.~\ref{fig:MRI_36_360}.

\subsection{Blazar component} 
\label{sec:blazar}

The division of AGN into radio-quiet and radio-loud is under discussion
as the arguments appeared both in favor of clearly bimodal distribution 
or against it, with the possible intermittency confusing the picture, 
and the origin of the phenomenon in unclear although the spin paradigm 
is attractive (see e.g. Sikora et al. 2007).

The support for some form of radio activity in radio-quiet objects is
strong.  Radio jets are present in numerous Seyfert 2 galaxies
(e.g. Pedlar et al. 1989; Falcke, Wilson \& Simpson 1998), and a less
extended radio emission is seen also in Seyfert 1 galaxies and Narrow
Line Seyfert 1 galaxies (Ulvestad \& Wilson 1989; Lal, Shastri \&
Gabuzda 2004; Doi et al. 2007). 
Some evidence of weak radio jets in a small number of RQQSOs
        has been reported from deep VLA imaging and related studies
(Miller, Rawlings \& Saunders 1993; 
Kellermann et al. 1994; Falcke et al. 1996a, 1996b).
There have also 
been some attempts at VLBA imaging at milliarcsecond resolution
 of the central engines. In this case, out
 of 12 radio-quiet QSOs, 8 of these sources
show strong evidence of a jet-producing central engine
       (Blundell \& Beasley 1998).
Deep VLBA imaging of 5 RQQSOs showed that only one source exhibits
 a two-sided radio jet and the other 4 are unresolved
       (Ulvestad, Antonucci \& Barvainis 2005).
Also, by radio monitoring with VLBA, relativistic jets were found in the
RQQSO in case of the source PG 1407$+$263 and in Seyfert 1 galaxy III Zw 2
 (Blundell, Beasley \& Bicknell 2003; Brunthaler et al. 2005).
One object --- 2MASX J0324+3410 ---
has all the properties of a standard Narrow Line Seyfert 1 galaxy and
still it is strong in radio, with $R \sim 100$, the radio image shows
a compact core and a well resolved one-sided jet, and radio emission
is accompanied by a possible detection of TeV gamma ray emission (Zhou
et al. 2007). It is possible that all X-ray power law
emission may actually come from a jet.  Such a possibility was
discussed mostly within the frame of galactic accreting black holes
(see e.g. Kylafis, Papadakis \& Reig 2007). However, the question is
open, and in particular, the correlation between the radio and the
X-ray emission is by no means simple (see e.g. Xue \& Cui 2007).

\subsubsection{lightcurves}

In the present paper we model the time dependence of the blazar
component using the cannonball model of the variability implemented by
Janiuk et al.  (2006) to model gamma-ray bursts. However, the model
can apply to all types of unstable jet-like outflow with a suitable
choice of the parameters.  The model is parametrized by the emission
radius, $r_0$, the Doppler parameter of the jet, $\Gamma$, the jet
opening, $\Theta_0$, the inclination of an observer, $i$ and the
number of burst events per unit of time, $N_{event}$.  We fix $r_0$ at
$10^{17}$ cm, which is generally appropriate for jet emission
(e.g. Hartman et al. 2001; Kataoka et al. 2007), we take $\Gamma = 10$
as representative for blazar jets. We fix $\Theta_0$ at 0.2 radian,
and $i$ at 0.6 radian as a representative example.

\begin{figure}
\epsfxsize=8.8cm 
\epsfbox{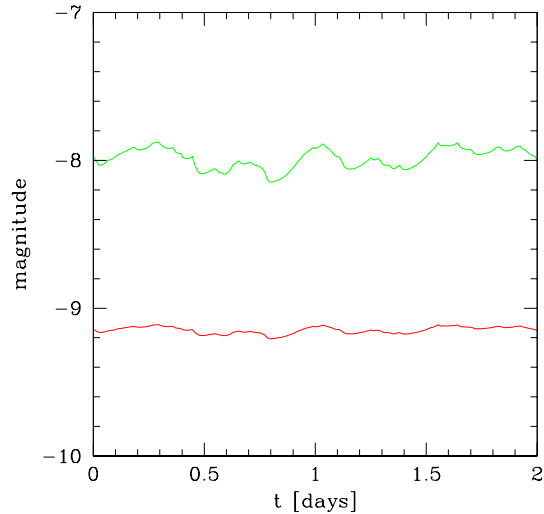}
\caption{The exemplary lightcurves (relative magnitudes) from the blazar component model without 
the disk contribution (upper curve) and with {\bf 80\% of stable disk}
contribution in V band (lower curve). Parameters: $r_0 = 10^{17}$ cm,
$\Theta_0$ = 0.2 rad, $\Gamma = 10$, $i = 0.6$ rad.
\label{fig:blazar_lightcurve}}
\end{figure}

The most important parameter model affecting the overall level of the
variability is $N_{event}$ per unit of time.  We adjust this parameter
describing the event rate to reproduce the variability properties of
pure blazar optical lightcurve.  Violent microvariability on blazars
can reach up to $\sim 0.08$ mag in one hour if spurious variability
detections are carefully avoided (Cellone, Romero
\& Araudo 2007), but such events are rare, and the variability is not 
always detected in a few hour observations. Therefore, in our model,
we fix $N_{event}$ at such a value (10~000 events per year) which
gives 50\% probability of detecting variability at $\psi = 2$ \% (see
eq.~\ref{eq:psi}) in 3h observation when no stable contribution to the
optical lightcurve is included. An exemplary lightcurve is shown in
Fig.~\ref{fig:blazar_lightcurve} where we show the pure blazar case as
well as the case when the stable accretion disk contributes 80 \% of
the total optical emission.

\subsubsection{broad band spectra models}

In order to model the average broad band spectrum we take two
components: standard stationary accretion disk emitting locally as a
black body, parametrized by the black hole mass and accretion rate,
and the blazar component with an arbitrary normalization. Since the
spectral shape of a blazar emission is well known to depend on the
luminosity, creating a blazar sequence, we consider two extreme cases
for the blazar shape.  The first case is a high luminosity
example. Specifically, we take two broad band spectra of the radio
loud quasar 3C 279, states P1 and P9 from Hartman et al. (2001).  For
this blazar component alone, the $\alpha_{ox}$ parameter is 1.17 and
1.27, correspondingly.  Any disk contribution will increase the
expected $\alpha_{ox}$.  As a second example, we take a lower
luminosity example of BL Lacs shown as LBL spectrum (BL Lac type
spectrum with peak at Low frequencies) in a review by Becker et
al. (2007).  We do not consider HBL (i.e. BL Lac type spectrum with
peak at High frequencies) since it is shifted too far into high energy
band and shows TeV emission.  The LBL blazar component alone is much
steeper, with $\alpha_{ox}$ equal to 1.65.

\section{Results}
\label{sec:results}

The three microvariability mechanisms implemented in a form of
numerical models of the lightcurves allow us to 
estimate the probability to observe an intra-night 
variability of an AGN within the frame of the three scenarios. 

\subsection{X-ray irradiation of an accretion disk}
\label{sect:X-ray_irrad_result}

\begin{figure}
\epsfxsize=8.8cm 
\epsfbox{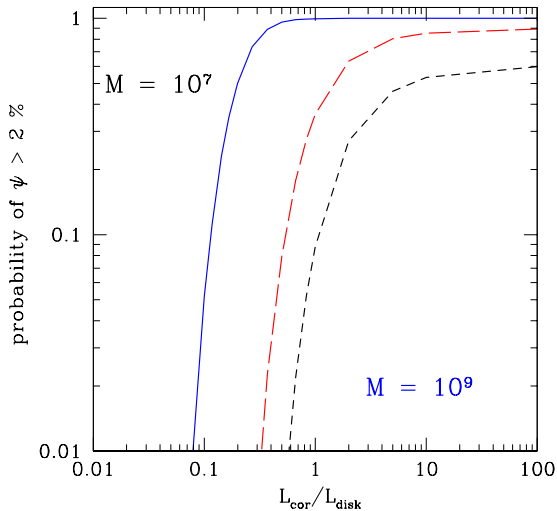}
\caption{Predicted probability of the microvariability at the level
exceeding two per cent for the X-ray irradiation model, as a function
of the corona to disk luminosity ratio, for three
values of the central black hole mass, $10^7$, $10^8$, and $10^9
M_{\odot}$, and the bolometric luminosities $10^{45}$, $10^{46}$ and
{\bf $10^{47}$} erg s$^{-1}$, correspondingly. 
\label{fig:irrad_probab1}}
\end{figure}

The predicted level of the optical variability strongly depends 
on the black hole mass and luminosity, as well as  
the X-ray coronal luminosity to the optical disk 
luminosity ratio, if the parameterization by Eq.~\ref{eq:break1} is
used to derive the X-ray lightcurve. The exemplary results, for the
Eddington ratio of $\sim 0.7$,
are shown in Fig.~\ref{fig:irrad_probab1}.

The model does not seem to account for the observed variability.
The INV is unlikely to be seen if the black hole mass is 
$\sim 10^8 M_{\odot}$ or more unless the X-ray emission is strong.

The problem is that the X-ray variability timescales are {\bf too
long, or equivalently, the X-ray variability amplitude at the required
timescale is not high enough to give the visible effect, when the
fraction of the energy dissipated in the corona is low.} The high
frequency breaks for the presented models are: $1.9 \times 10^{-4}$
Hz, $1.4 \times 10^{-5}$ Hz and $1.1 \times 10^{-6}$ Hz,
correspondingly. Thus for higher mass objects {\large amplitude}
variability is on the timescales of a week or longer.  The smaller
black hole mass ($\sim 10^7 M_{\odot}$) predicts significant INV (see
Fig.~\ref{fig:irrad_probab1}) but such a mass is unlikely to be
appropriate for quasars.

\begin{figure}
\epsfxsize=8.8cm 
\epsfbox{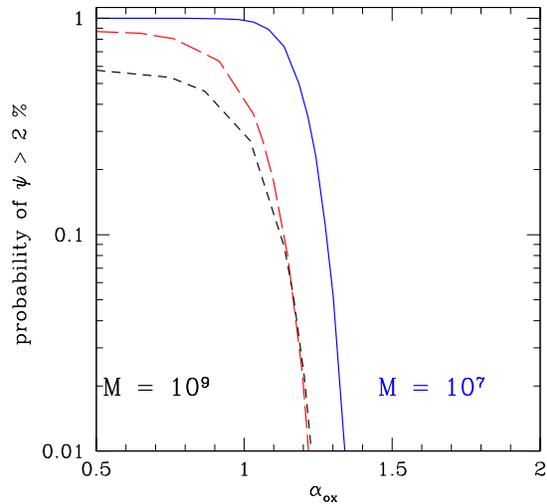}
\caption{Predicted probability of the microvariability at the level
exceeding two per cent for the X-ray irradiation model, as a function
of the broad band spectral slope, for three
models shown in Fig.~\ref{fig:irrad_probab1}.
\label{fig:irrad_probab2}}
\end{figure}

{\bf We cannot enhance the variability by introducing a stronger
irradiation since a high} fraction of the energy dissipated in the
corona is not compatible with the broad band spectral slope. In
Fig.~\ref{fig:irrad_probab2} we show the predicted variability level
as a function of $\alpha_{ox}$. For larger black hole mass ($10^8 -
10^9 M_{\odot}$), more appropriate for quasars, and steep spectra, the
INV is always very low, and for $\alpha_{ox} \sim 1.5$ the probability
drops below $10^{-6}$.

In these considerations, however, we neglected the issue {\it where} in the
disk the reprocessing occurs since we simply assumed that the typical
dilution in the observed optical band is the same as the overall ratio
of the disk to corona average emissivity. 

Since this effect is potentially
important we separately consider the effect how the sudden
irradiation by an X-ray flare changes the disk flux at a specific
wavelength.  As an extreme example, we
choose a situation when only a single flare forms above an accretion
disk. A flare contains a specified fraction of the
total bolometric luminosity of the source (i.e. all the corona
luminosity), it is located at a given
radius, and the size of the hot spot beneath the flare
(i.e. reprocessing area) is specified as a fixed fraction of the
radius (physically, it is given by the flare height).

\begin{figure}
\epsfxsize=8.8cm 
\epsfbox{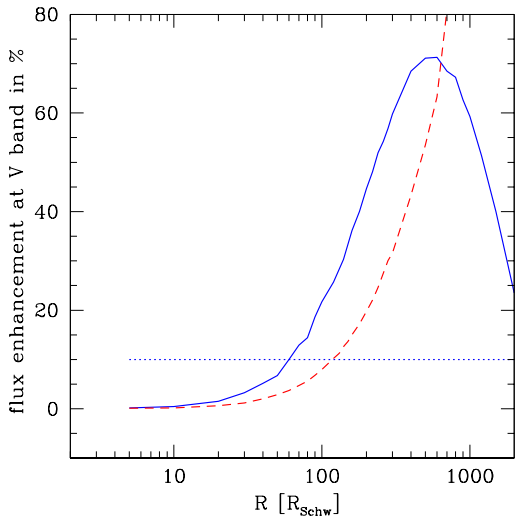}
\caption{The relative enhancement in the radiation flux in V band as a
  function of the location of a single irradiating flare (continuous
  line) for $M = 10^9 M_{\odot}$, $L = 10^{47}$ erg s $^{-1}$,
  $L_{cor}/L_{disk} = 0.1$ and the spot size 0.1 of its position
  radius. Dotted line marks the adopted disk/corona ratio. Dashed line
  shows the enhancement for a stronger corona $L_{cor}/L_{disk} = 0.2$
  and more compact flare, of the size of 0.01 of its position radius.
\label{fig:local_reprocessing}}
\end{figure}

A specific example for a large mass black hole is shown in 
Fig.~\ref{fig:local_reprocessing}. The enhancement is less than the
simple corona to disk ratio for radii smaller than $50 R_{Schw}$ but
then it increases rapidly. This means that if extremely strong single 
flares are occasionally localized
at $400 R_{Schw}$ the enhancement factor is almost 80 \% in
V band, so the {\it local} disk to corona ratio is is not 0.1, but
0.8! With such a small dilution, the probability to find an intraday
variations is 5.2 \% while on average  the overall slope
$\alpha_{ox}$ may be the same as
for $L_{cor}/L_{disk} = 0.1$, i.e. 1.52, quite consistent with the
data.

Therefore, if very strong flares are likely to appear at large disk
radii, the X-ray irradiation may be the cause of the observed
intranight variability. 

\begin{figure}
\epsfxsize=8.8cm 
\epsfbox{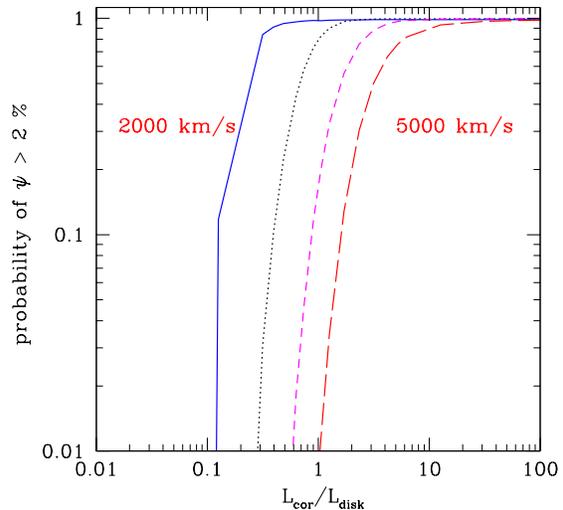}
\caption{Predicted probability of the microvariability at the level
exceeding two per cent for the X-ray irradiation model, as a function
of the corona to disk luminosity ratio, for four 
values of FWHM of H$\beta$ line {\bf (2000, 3000, 4000 and 
5000 km s$^{-1}$)}. 
\label{fig:irrad_probab_vel}}
\end{figure}

If the X-ray lightcurve is generated using the Eq.~\ref{eq:break2},
the chances to have the INV due to X-ray irradiation in our sample are
higher. Although the formula ~\ref{eq:break2} gives a relatively large value of
the high frequency break, $7.8 \times 10^{-6}$ Hz, for the line width 
of 3000 km s$^{-1}$,
possibly representative for our sample, and in that case the INV is
not expected. However, for objects with very
narrow lines, like Mkn 1014, the irradiation scenario seems quite
promising. In this case we have no direct
relation to the broad band parameterization of the spectral shape.

\subsection{Disk instability}

\subsubsection{Radiation pressure instability}

\begin{figure}
\epsfxsize=8.8cm 
\epsfbox{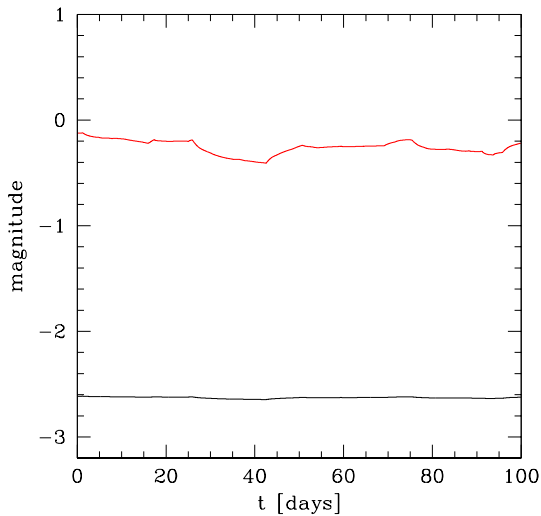}
\caption{An exemplary short timescale lightcurve of the disk with
  corona during the disk maximum luminosity during the outburst, for
  the same parameters as in Fig.~\ref{fig:outburst}. 
\label{fig:curve_mag_sqrt}}
\end{figure}

We used the $\sqrt{P_{\rm gas} P_{\rm tot}}$ viscosity law instead of
$P_{\rm tot}$ law of Shakura \& Sunyaev (1973) with the aim to obtain
lower amplitude and shorter outbursts. Indeed, this new prescription
gives outbursts lasting by a factor of a few hundred shorter than the
standard law. However, this is still far too long to model INV. The
exemplary lightcurve shown in Fig.~\ref{fig:outburst} was obtained for
a black hole mass $10^{8} M_{\odot}$ 
and a viscosity parameter $\alpha_{\rm disk} = 0.01$ . The outburst
last thousands of years.

The computations performed without a corona did not show any short
timescale additional thermal oscillations during the evolution, in
agreement with previous computations (e.g. Merloni \& Nayakshin
2006). Of course a fast variability reappears if the corona is included
(e.g. Mayer \& Pringle 2006, see also Fig.~\ref{fig:curve_mag_sqrt})
but in this case the problem reduces to the variability model
discussed in Sect.~\ref{sect:X-ray_irrad_result}. For this specific
model, the probability of the INV is below $10^{-3}$. Although there
is some coupling in theoretical models between the presence or absence
of an outburst and the coronal variability (through the change in the
disk thickness; Mayer \& Pringle 2006; Janiuk \& Czerny 2007) it
cannot modify the disk/corona system behaviour in a short timescale.

\subsubsection{magnetorotational instability}
 
\begin{figure}
\epsfxsize=8.8cm 
\epsfbox{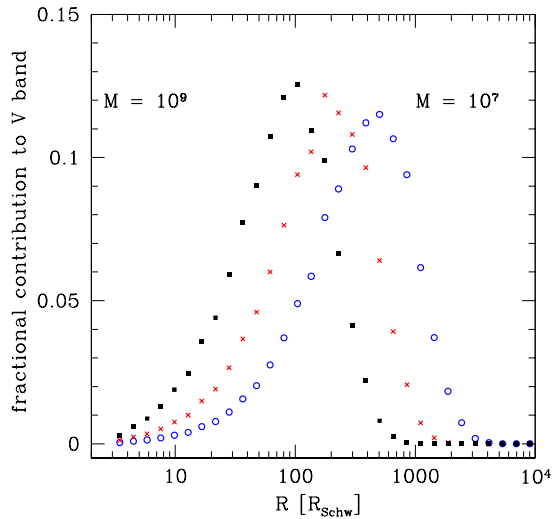}
\caption{The relative contribution to the V band as a function of
  radius for three disk models:$M = 10^7$, $10^8$ and $10^9
  M_{\odot}$, and the bolometric luminosities $10^{45}$, $10^{46}$ and
  $10^{47}$ erg s$^{-1}$, correspondingly.
\label{fig:udzialy_w_V}}
\end{figure}

The size of a magnetic cell is not known but there are natural
contraints of its value. The cell should be smaller than the local
thickness of the accretion disk, $H_d$. It also should be high enough
to account for the disk viscosity. Since the coefficient
$\alpha_{disk}$ (Shakura \& Sunyaev 1973) can be interpreted as a
ratio of the turbulent velocity to the sound speed multiplied by the
ratio of the cell size to the disk thickness, the minimum size of the
cell is $\alpha H_d$. The viscosity coefficient $\alpha$ in quasars
estimated from large amplitude variability is of order of 0.02
(Siemiginowska \& Czerny 1989; Starling et al. 2004), consistent with
numerical MHD simulations. Thus, at a single disk ring of the radius
$r$, and the radial width, $H_d$, the number of cells is
\begin{equation}
  {2 \pi r \over H_d}  < N_{cell} < 10^5 {2 \pi r \over H_d}.
\end{equation}
The value $H_d/r = 0.3$ is an adequate representation even of the high
Eddington ratio objects (see e.g. Fig. 3 of Loska et al. 2004).

The total number of cells in the whole disk is large but not the whole
disk participates in providing the luminosity at a given wavelength.

We constrain that by considering a standard stationary Shakura-Sunyaev
disk for a given black hole mass and bolometric luminosity and we
determine the fractional contribution of the disk emission to the V
band as a function of radius, as shown in Fig.~\ref{fig:udzialy_w_V}. 
We finally divide the disk into rings of
the widths $H_d(r)$ increasing with radius, we model each ring as
described above taking the most favorable condition for
microvariability, $N_{cell} = 2 \pi
r/H_d = 20$.

\begin{figure}
\epsfxsize=8.8cm 
\epsfbox{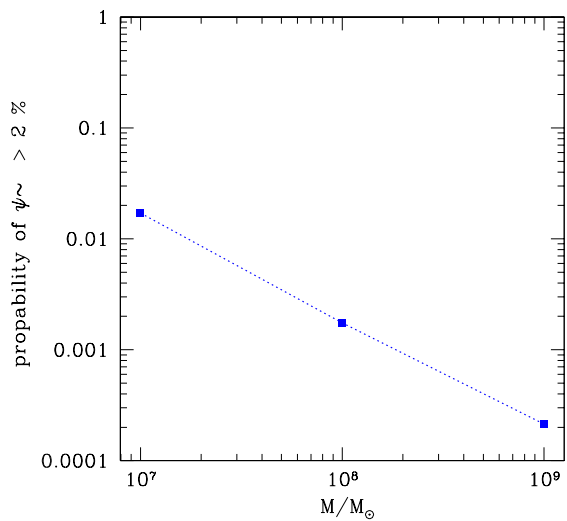}
\caption{The probability of the INV in V band caused by MRI
 variability for three disk models:$M = 10^7$, $10^8$ and $10^9
  M_{\odot}$, and the bolometric luminosities $10^{45}$, $10^{46}$ and
  $10^{47}$ erg s$^{-1}$, correspondingly.
\label{fig:val_trend_MRI_Vband}}
\end{figure}

The results are shown in Fig.~\ref{fig:val_trend_MRI_Vband}. The
probability of the INV variability in more massive objects is very
low. It is caused by the increase of the timescales with the black
hole mass which is not quite compensated by the decrease in the radius
contributing most strongly to the  V band (100 $R_{Schw}$ for $10^9
M_{\odot}$ in comparison with 400 $R_{Schw}$ for $10^7
M_{\odot}$, see Fig.~\ref{fig:udzialy_w_V}).

\subsection{Blazar component}

The predicted probability of an INV depends on the disk contribution
at the monitored wavelength. Since the model was adjusted to have 50\%
probability of the INV for a pure blazar component in 3 h
observations, and our modeling of radio quiet sources assumes 5 h
observations the probability of the INV starts at 84\% for
negligible disk and drops to moderate 20\% when the disk to the blazar
luminosity
ratio reaches 2.7. This is interesting in the context of the
equivalent width of H$\beta$ line. The observed equivalent width of
H$\beta$ line in four sources from Table~\ref{tab:xrays} is on average 
$63 \pm 20$
\AA~, comparable to the average of 62.4 \AA~ in the whole quasar
sample of (Forster et al. 2001). Therefore we do not seem to see
a decrease of the line width 
due to the dilution by the blazar component. Therefore, the 
blazar contribution at H$\beta$ wavelength should be low. 

\begin{figure}
\epsfxsize=8.8cm
\epsfbox{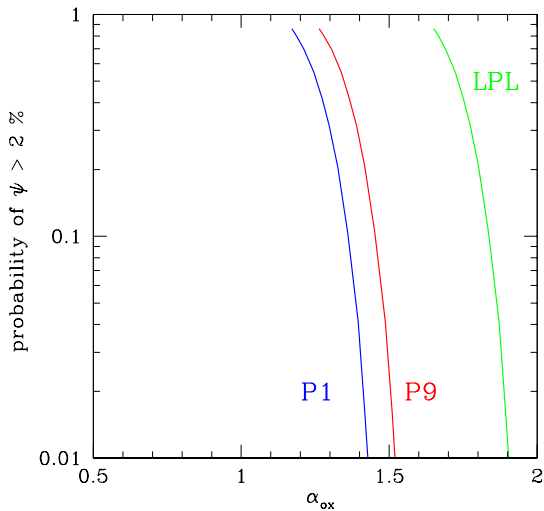}
\caption{The probability of the intra-night variability for a black hole mass 
$10^8_{\odot}$ as a function of the $\alpha_{ox}$ for the two blazar shapes 
(state P1 and P9 of 3C 273 from Hartman et al. 2001) and the Low frequency peak 
BL type shape (state LBL) after Becker (2007). 
\label{fig:blazar2}}
\end{figure}

The relation between the INV and the broad band spectral shape depends
on the spectral shape of the blazar component.  The results for the
three specific cases of our choice are shown in Fig.~\ref{fig:blazar2}
where we show the INV probability as a function of the disk
contribution translated to the observed parameter $\alpha_{ox}$. In
all cases the predicted variability level was still significant if the
$\alpha_{ox}$ approached 1.5 for the black hole mass $10^8
M_{\odot}$. The range of slopes for each of the specific shapes was
too narrow to reproduce the whole sample given in
Table~\ref{tab:xrays} but we cannot expect a single universal blazar
spectral shape to work if even the same source vary in its spectral
shape significantly. If the blazar spectral shape is allowed to vary
from source to source, as it is the case in true blazars, the blazar
component model has the flexibility enough to reproduce the sample
properties. 10 - 20 \% variability was obtained with the blazar
component contribution of 10 - 8.5 \% which is consistent with no
significant influence on the h$\beta$ equivalent width.

\section{Discussion}
\label{sec:diss}

In the present paper we collected a sample of radio-quiet objects with
confirmed microvariability in the optical band (i.e. intra-night
variability) for which also X-ray data were available. As radio-quiet,
we treated sources with the standard loudness parameter R $< $10. There
were only 10 of such objects but even such a small sample allowed for
an interesting analysis.

The broad band spectra of variable objects did not differ
significantly from objects with no detection of variability: the median
value of $\alpha_{ox}$ parameter was 1.47 in variable objects, and
1.52 in non-variable objects. {\bf Also derived values of equivalent width 
of H$\beta$ line did not differ significantly from the typical values
in non-variable objects.  }
 However, it was measured only for four
objects in our sample so at present we cannot claim this effect to be
proven.

The main goal of our paper was to test three possible scenarios of the
optical microvariability: (i) reprocessing of the variable X-ray
emission (ii) accretion disk instability (iii) weak blazar component
in radio-quiet sources.  For these three scenarios we formulated
specific models which allowed to obtain the lightcurves and to analyse
the probability of the detection of microvariability in a single
observation lasting 5 h.

The first scenario has difficulties to account for the variability
which is rapid enough to show up in a form of intra-night erratic
variations,  {\bf without being  in conflict with a broad band spectral
shape}.  The result seems to be generic, as the typical rms value of
the X-ray variability in 1 d timescale is 3\% (Markowitz \& Edelson
2004) for black hole masses of order of $10^8 M_{\odot}$, and if X-ray
flux is by a factor 10 lower than the optical disk emission, as
typical in quasars, the optical amplitude of variations is below the
detection threshold. {\bf A larger ratio of the X-ray flux to the
optical disk emission is in turn inconsistent with broad band quasar
spectra}. However, for objects closer in their nature to Narrow Line
Seyfert 1 this variability may be enhanced, and the lightcurves
obtained assuming FWHM of H$\beta$ line 2000 km s$^{-1}$ show INV
pattern.

There is also another possibility of the variability enhancement,
namely extremely bright single coronal flares, containing significant
fraction of the bolometric luminosity but located at large disk radii
also lead to large variability in the visual band. Such a flare heats
up the previously cold disk surface to high temperatures and the extra
flux appears right in the visual band. From the
theoretical point of view, such flares are hardly expected, as the amount of
energy available at large radii is small. On the other hand, such
flares were occasionally reported and modeled in the context of X-ray
reprocessing (e.g. Goosmann et al. 2004). In this case the smearing
effect due to the light travel time must be considered, with the net
effect depending additionally on the inclination of an observer.

The disk instabilities were studied taking into account two
mechanisms. First we studied the radiation pressure instability. 
 We used the viscosity prescription
advocated as the most promising by Merloni \& Nayakshin (2006) in
comparison to the observed properties of the accreting black
holes. This model, however, gives only long-lasting outbursts not
suitable for microvariability since no short timescale thermal
pulsations accompany this instability, if the X-ray coronal emission
is neglected. If the coronal emission is modeled, the situation
practically reduces to the previous scenario of variable X-ray
irradiation. Therefore we next deviced a crude model of much faster, 
magneto-rotational instability. However, even in this case the
variations are not fast enough to account for a rapid flux
changes. Since the optical emission comes from the radius of $\sim 100
R_{Schw}$ or more, this is not surprising. More advanced simulations
are consistent with our results. 
Local magneto-rotational instability coupled with
radiation pressure instability is likely to lead to thermal erratic
oscillations (Turner 2004) accompanied by a change in the local disk
luminosity by a factor of a few but on the timescales of order of
a few months for a $10^8 M_{\odot}$ object at 100 $R_{Schw}$. 

The third scenario is easily acceptable. Although we tested our blazar
component model for a fixed values of parameters (i.e. viewing angle,
number of events, Lorentz factor, emission radius, etc.), this set of
parameters represents well the observed blazar emission. In case of
the radio quiet versus radio loud objects, the modeled pulse profiles
depend mainly on whether the viewing angle is smaller or larger than
the jet opening angle, and not that much on its particular
value. Therefore we conclude that our calculations are representative
for a sample of objects, which of course may have various
inclinations.

{\bf The blazar contribution required to explain the INV was about 20\%,
and it was not in conflict with the broad band spectra.} The broad
range of the $\alpha_{ox}$ within the sample can be reproduced only
with a range of the shapes of the blazar component but the blazars
themselves also show a broad range of spectra. Our result thus comes
in line with the conclusion of Carini et al. (2007) that the
probability of microvariability is a smoothly decreasing function of
the radio loudness indicating that the same mechanism operates in
blazars and in the radio-quiet sources.

Our result may support the view that even radio-quiet sources
produce a beamed blazar-like emission but in radio-quiet objects this
emission is weaker.  
On the other hand, longer timescale variations of the radio-quiet sources in
the optical band are apparently of a different nature, mostly
reprocessing (see e.g. Enya et al. 2002), and possibly also disk
instabilities (e.g. Czerny et al. 1999).


\section*{Acknowledgments}

{\bf We are grateful to the anonymous refreree who pressed us to have
a better insight into the studied phenomenon}.  This research has made
use of data obtained through the HEASARC Online Service, provided by
NASA/Goddard Space Flight Center and of the data from the Sloan
Digital Sky Survey (http://www.sdss.org). This work was supported in
part by grant 1P03D00829 of the Polish State Committee for Scientific
Research and the Polish Astroparticle Network
621/E-78/SN-0068/2007. Partial support for this work was provided by
the National Aeronautics and Space Administration through Chandra
Award Number GO5-6113X and under the contract NAS8-39073.


\subsection{Appendix: Archival X-ray data}

{\bf PG0043+039}

A broad absorption line (BAL) quasar PG0043+039 (z=0.385) was not
detected in the {\it ASCA} pointed observation on Dec.21 1996.
Gallagher et al. (1999) give a 3 $\sigma$ upper limit of $2 \times
10^{13}$ erg~s$^{-1}$~cm$^{-2}$ for 2-20 keV flux obtained for the
adopted photon index $\Gamma = 2.0$.  BAL quasars are known to have a
weak X-ray emission due to an intrinsic absorption and the {\it ASCA}
non-detection gives a typical lower limit to the absorber's column
density of $N_H > 4\times 10^23$~cm$^{-2}$.

The quasar was also undetected in the 30~ksec XMM-{\it Newton}
observation on June 15, 2005.  We extracted the data from the XMM-{\it
Newton} archive and processed using the SAS version 7.0.0.  We
determined an upper limit for the count rate at the position of the
source to be 0.001 cts/sec in 0.3-8 keV energy band. This corresponds
to the 3$\sigma$ upper limit for the flux density at 1 keV of $8.6
\times 10^{16}$ erg s$^{-1}$ cm$^{-2}$ keV$^{-1}$, for the assumed
photon index $\Gamma = 1.7$ and Galactic absorption $N_{\rm H} =3.2
\times 10^{20}$ cm$^{-2}$. Non-detection in XMM-{\it Newton} observation
at this flux level indicate a column density large than $N_H >
10^{25}~cm^{-2}$.

\begin{figure}
\epsfxsize=8.8cm 
\epsfbox{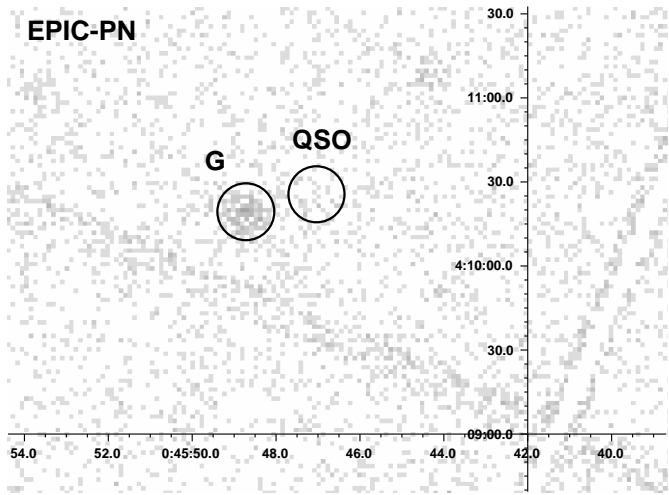}
\caption{PG0043+039 in XMM-{\it Newton}. The EPIC/PN image in 
0.5-10~keV energy range.  A nearby galaxy is marked as G and
PG0043+039 is marked as QSO and circular regions with 10 arcsec radii.
\label{fig:pg0043_map}}
\end{figure}

Figure~\ref{fig:pg0043_map} shows the XMM-{\it Newton} field of view
of EPIC/PN with the location of the quasar and a nearby galaxy marked
on the image. The quasar was not detected.


\medskip
\noindent
{\bf PG0832+251}


The quasar PG0832+251 (z=0.331) was in the field of view during one of
the ROSAT PSPC observations in May 1992. The exposure time was 2792
sec. We obtained the 3$\sigma$ upper limit to the count rate, $1.42
\times 10^{-2}$ cts/sec, and we converted this limit to 0.5 -2 keV
flux of $1.3 \times 10^{-13}$ erg s$^{-1}$ cm$^{-2}$ using
PIMMS\footnote{http://heasarc.gsfc.nasa.gov/Tools/w3pimms.html}(Mukai
1993) and assuming the Galactic column density of $3.62 \times
10^{20}$ cm$^{-2}$, and a photon index $\Gamma = 1.7$.

\medskip
\noindent
{\bf 1422+424}

The source 1422+424 was observed with {\it Chandra} X-ray Observatory
on Sep.25, 2002 for about 6~ksec (Risaliti et al. 2003). The source is
relatively X-ray weak, but not significantly reddened, with the photon
index $\Gamma$ equal to $2.26 \pm 0.08$.  The X-ray flux is $5.2\times
10^{-14}$ erg s$^{-1}$ cm$^{-2}$ in 0.5-2 keV band,and $4.27 \times
10^{-14}$ erg s$^{-1}$ cm$^{-2}$ in 2 - 10 keV band (Risaliti et
al. 2003, Risaliti et al, in preparation private communication).

\label{lastpage}
\end{document}